\documentclass[apjl]{emulateapj}

\usepackage{url}
\usepackage{color}
\usepackage{bm}

\shorttitle{Millimeter Polarized Emission and Faraday Screen of 3C~84}
\shortauthors{Nagai et al.}

\begin{document}


\title{Enhanced Polarized Emission from the One-Parsec-scale Hotspot of 3C~84 as a Result of the Interaction with Clumpy Ambient Medium}


\author{H. Nagai\altaffilmark{1,2}, Y. Fujita\altaffilmark{3}, M. Nakamura\altaffilmark{4}, M. Orienti\altaffilmark{5,7}, M. Kino\altaffilmark{6,1}, K. Asada\altaffilmark{4}, G. Giovannini\altaffilmark{5,7}}



\altaffiltext{1}{National Astronomical Observatory of Japan, Osawa 2-21-1, Mitaka, Tokyo 181-8588, Japan}\email{hiroshi.nagai@nao.ac.jp}
\altaffiltext{2}{The Graduate University for Advanced Studies (SOUKENDAI), Osawa 2-21-1, Mitaka, Tokyo 181-8588, Japan}
\altaffiltext{3}{Theoretical Astrophysics, Department of Earth and Space Science, Graduate School of Science, Osaka University, 1-1 Machikaneyama-cho, Toyonaka, Osaka 560-0043, Japan}
\altaffiltext{4}{The Academia Sinica Institute of Astronomy and Astrophysics, AS/NTU. No.1, Sec. 4, Roosevelt Rd, Taipei 10617, Taiwan, R.O.C.}
\altaffiltext{5}{Instituto di Radioastronomia, Via P. Gobetti, 101 40129 Bologna, Italy}
\altaffiltext{6}{Kogakuin University, Academic Support Center, 2665-1 Nakano, Hachioji, Tokyo 192-0015, Japan} 
\altaffiltext{7}{Dipartimento di Fisica e Astronomia, Universita' di Bologna, via Gobetti 83b, I 40129 Bologna, Italy}

\begin{abstract}
We present Very Long Baseline Array polarimetric observations of the innermost jet of 3C~84 (NGC~1275) at 43~GHz.  A significant polarized emission is detected at the hotspot of the innermost re-started jet, which is located ~1 pc south from the radio core.  While the previous report presented a hotspot at the southern end of the western limb, the hotspot location has been moved to the southern end of the eastern limb.  Faraday rotation is detected within an entire bandwidth of the 43-GHz band.  The measured rotation measure (RM) is at most (6.3$\pm$1.9)$\times10^{5}$~rad~m$^{-2}$ and might be slightly time variable on the timescale of a month by a factor of a few.  Our measured RM and the RM previously reported by the CARMA and SMA observations cannot be consistently explained by the spherical accretion flow with a power-law profile.  We propose that a clumpy/inhomogeneous ambient medium is responsible for the observed rotation measure.  Using equipartition magnetic field, we derive the electron density of $2\times10^{4}$~cm$^{-3}$.  Such an electron density is consistent with the cloud of narrow line emission region around the central engine.  We also discuss the magnetic field configuration from black hole scale to pc scale and the origin of low polarization.   
\end{abstract}		


\section{Introduction}
3C~84 is associated with the giant elliptical galaxy NGC~1275 (z=0.0176), whose nuclear emission is classified as a Seyfert 1.5/LINER (Sosa-Brito et al. 2001), centered at the Perseus cluster.  The total luminosity is $4\times10^{44}$~erg~s$^{-1}$ (Levinson et al.1995).  This luminosity is about 0.4\% of the Eddington luminosity for a black hole mass of $8\times10^{8} M_{\odot}$ (Scharwachter et al. 2013).  The radio luminosity of this source is $3\times10^{24}$~W~Hz$^{-1}$~sr$^{-1}$ at 178~MHz, which is classified as a Fanaroff-Riley I radio source (Fanaroff \& Riley 1974).  The radio morphology is quite complex.  Multiple radio lobes are found on different angular scales, suggesting intermittent radio jet activities.  The most recent jet activity started in $\sim$2005 (Nagai et al. 2010).  The restarted jet extends up to $\sim1$~pc to the south from the core and a hotspot and lobe-like structure are formed at the southern end (e.g., Nagai et al. 2014).  The counter jet has also recently been discovered (Fujita \& Nagai 2017).  Based on the jet-counter jet ratio, Fujita and Nagai (2017) estimated that the jet forms an angle of $65^{\circ}\pm15^{\circ}$ with the line of sight, which is similar to the one for the jet associated with the previous episode of activity (e.g., Asada et al. 2006).  The total radio flux density is about 40~Jy at 10~GHz\footnote{F-GAMMA program: \url{http://www3.mpifr-bonn.mpg.de/div/vlbi/fgamma/fgamma.html}}.  Most of the radio emissions originate in the innermost jet\footnote{MOJAVE program: \url{http://www.physics.purdue.edu/MOJAVE/index.html}}.

NGC~1275 has been the subject of extensive studies of the circumnuclear gas properties in connection with the mass accretion onto the super massive black hole (SMBH).  NGC~1275 is known to have a large reservoir of cold molecular gas (Salome et al. 2006; Lim et al. 2008) in contrast to other brightest cluster galaxies (BCGs) such as M87 \citep{Tan2008, Perlman2007}.  The total amount of molecular gas mass ($M_{\rm gas}$) is $M_{\rm gas}\simeq10^{10}M_{\odot}$.  A large fraction of this molecular gas is located within the central 1~kpc ($M_{\rm gas}\simeq10^{9}M_{\odot}$: Lim et al. 2008).  In the inner 50~pc, the circumnuclear disk (CND) is resolved by the warm H$_{2}$ and ionized [Fe II] lines, both in morphology and kinematics with the Gemini North Telescope (Sch{\"a}rwachter et al. 2013).  It was suggested from the observed velocity dispersion that the H$_{2}$ emission traces the outer region of the disk which is likely to form a toroid while the [Fe II] line traces the inner region of the disk illuminated by the ionizing photons from the active galactic nucleus (AGN).  The inner ionized part is possibly associated with the ``silhouette" disk, which was identified by the free-free absorption (FFA) of background synchrotron emission from the counter jet by VLBI observations (Romney et al. 1995; Walker et al 2000).  

The intermittent jet activity of 3C~84 indirectly predicts that the accretion flow is strongly inhomogeneous.  Numerical simulations of giant elliptical galaxies suggest that the mass accretion is dominated by chaotic cold accretion within the inner kpc (Gaspari et al. 2013).  The simulations also predict that the chaotic cold accretion leads to a deflection of jets and strong
variation in the AGN luminosity.  Such a jet deflection and luminosity change are indeed observed in NGC~1275/3C~84 in radio, X-ray, and $\gamma$-ray bands (Nagai et al. 2010; Dutson et al. 2014; Fabian et al. 2015).  Fujita \& Nagai (2017) measured the opacity of free-free absorption ($\tau_{\rm ff}$) toward the counter jet component on the central $\sim1$~pc region and found $\tau_{\rm ff}\propto\nu^{-0.6}$, which is different from the one for a uniform density of $\tau_{\rm ff}\propto\nu^{-2}$.  They argued that the absorbing medium is highly inhomogeneous and that it consists of regions of $\tau_{\rm ff}\ll 1$ and $\tau_{\rm ff}\gg 1$.

Recently, VLBA images at 43 GHz from the Boston University Blazar Program\footnote{\url{http://www.bu.edu/blazars/VLBAproject.html}} have pointed out an enhancement of the polarized emission at the hotspot of the innermost jet as well as the abrupt change in its position.  At the beginning of 2013, the hotspot was located along the western limb \citep{Nagai2014}.  This situation was not changed until the middle of 2015.  After that, the hotspot on the western limb became less obvious, and the hotspot appeared at the eastern limb until the middle of 2016, accompanying with the polarized emission.  In more recent data, the hotspot structure has even become distorted.  These behaviors suggest a strong interaction between the jet and ambient medium.  The enhancement of the polarized emission is an important tool to probe the ambient medium on pc scales.  In this paper, we report the analysis of the total and polarized intensities in five epochs between 2015 December and 2016 April as well as Faraday rotation measure (RM) along the line of sight toward the polarized emission, which can constrain the electron density of the ambient medium.  Throughout this paper, we use $H_{0}$=70.5, $\Omega_{\rm M}=0.27$, and $\Omega_{\Lambda}=0.73$.  At the 3C 84 distance, 1~mas corresponds to 0.344~pc. 

\section{DATA}
We used the calibrated VLBA archival data taken as the part of Boston University Blazar Program.  The observations were done on 2015 December 05, 2016 January 01, 2016 January 31, 2016 March 18, and 2016 April 22 at 43~GHz with ten VLBA stations.  The data consist of four intermediate frequencies (IFs) with a 64-MHz bandwidth for each IF.  The central frequencies of each band are 43.008, 43.087, 43.151, and 43.215~GHz for IF 1, 2, 3, and 4, respectively.  The total bandwidth is 256 MHz per polarization. Both right-hand (R) and left-hand (L) circular polarizations were received, and RR, LL, RL, and LR correlations were obtained.  The calibrations were done in the same manner as describe in \cite{Jorstad2005}.  The instrumental polarizations (D-terms) were derived by averaging values over thirteen sources observed in each epoch in each IF using AIPS task LPCAL.  The absolute electric vector position angle (EVPA) calibration was done with the D-term method \citep{Gomez2002} and using jet features with stable EVPAs in several quasars.

\begin{figure*}
\begin{center}
\begin{tabular}{cc}
 \begin{minipage}{0.5\hsize}
  \begin{center}
   \includegraphics[width=8cm]{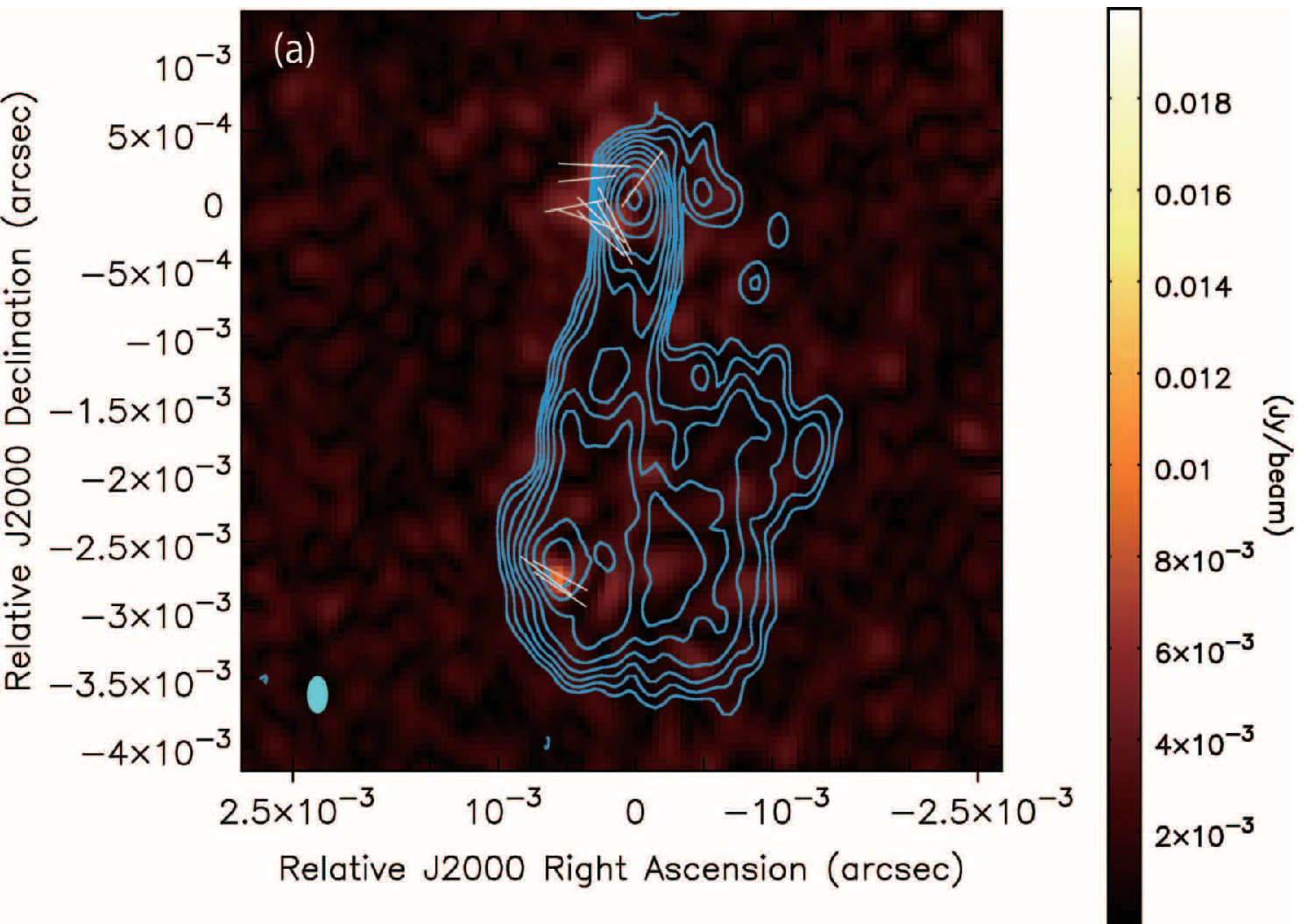}
  \end{center}
 \end{minipage}
 \begin{minipage}{0.5\hsize}
  \begin{center}
   \includegraphics[width=8cm]{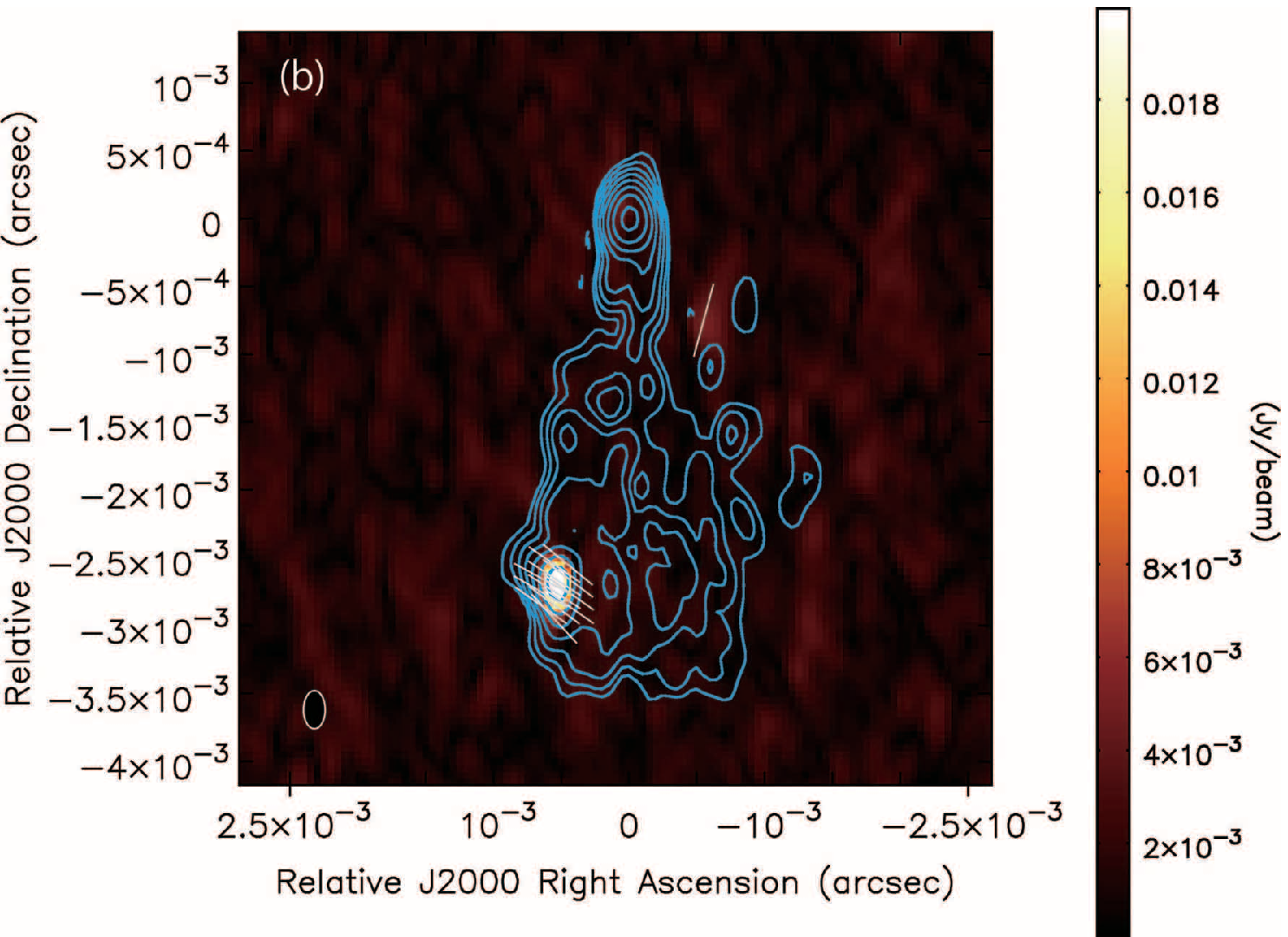}
  \end{center}
 \end{minipage}\\
\begin{minipage}{0.5\hsize}
  \begin{center}
   \includegraphics[width=8cm]{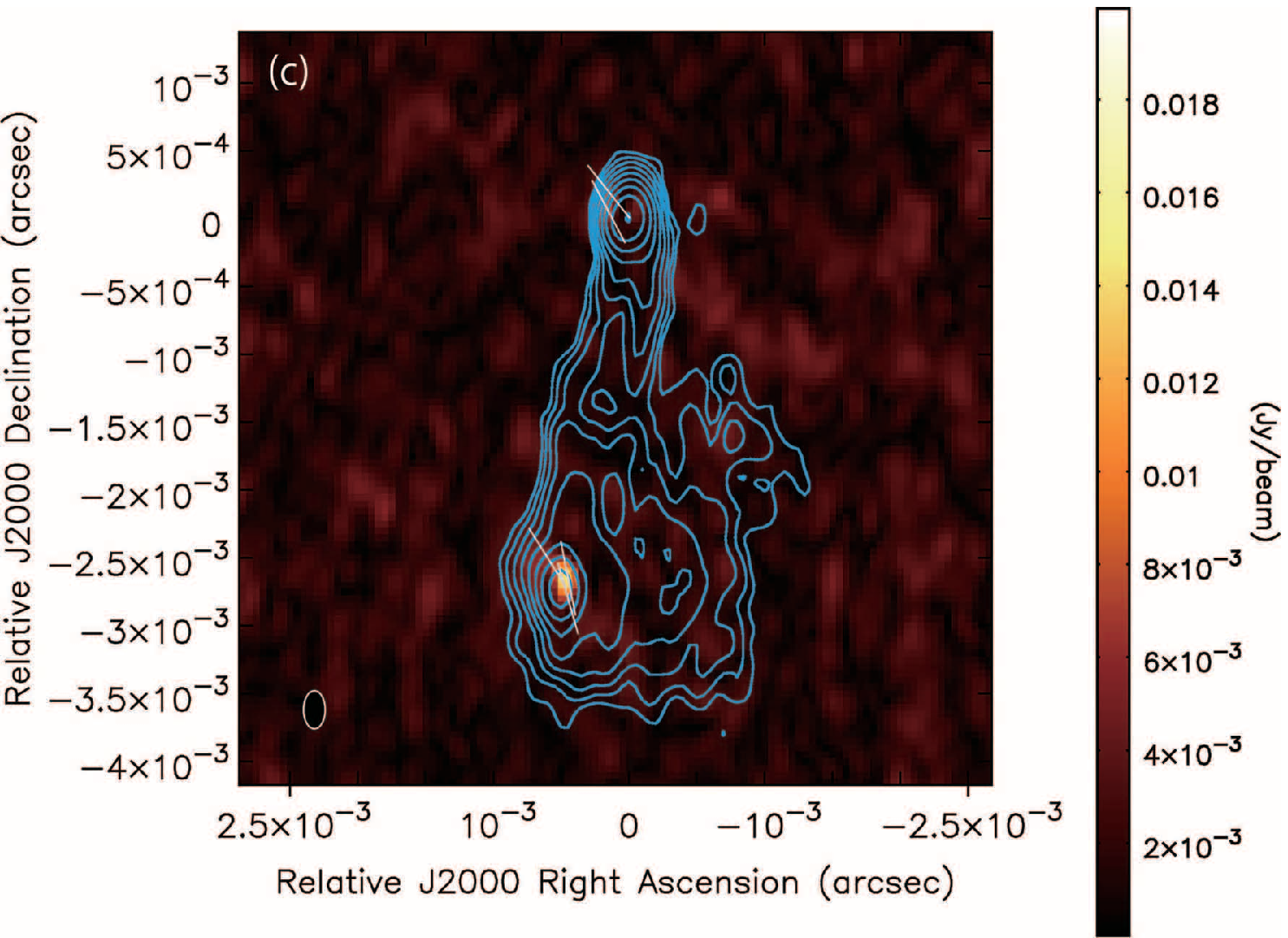}
  \end{center}
 \end{minipage} 
\begin{minipage}{0.5\hsize}
  \begin{center}
   \includegraphics[width=8cm]{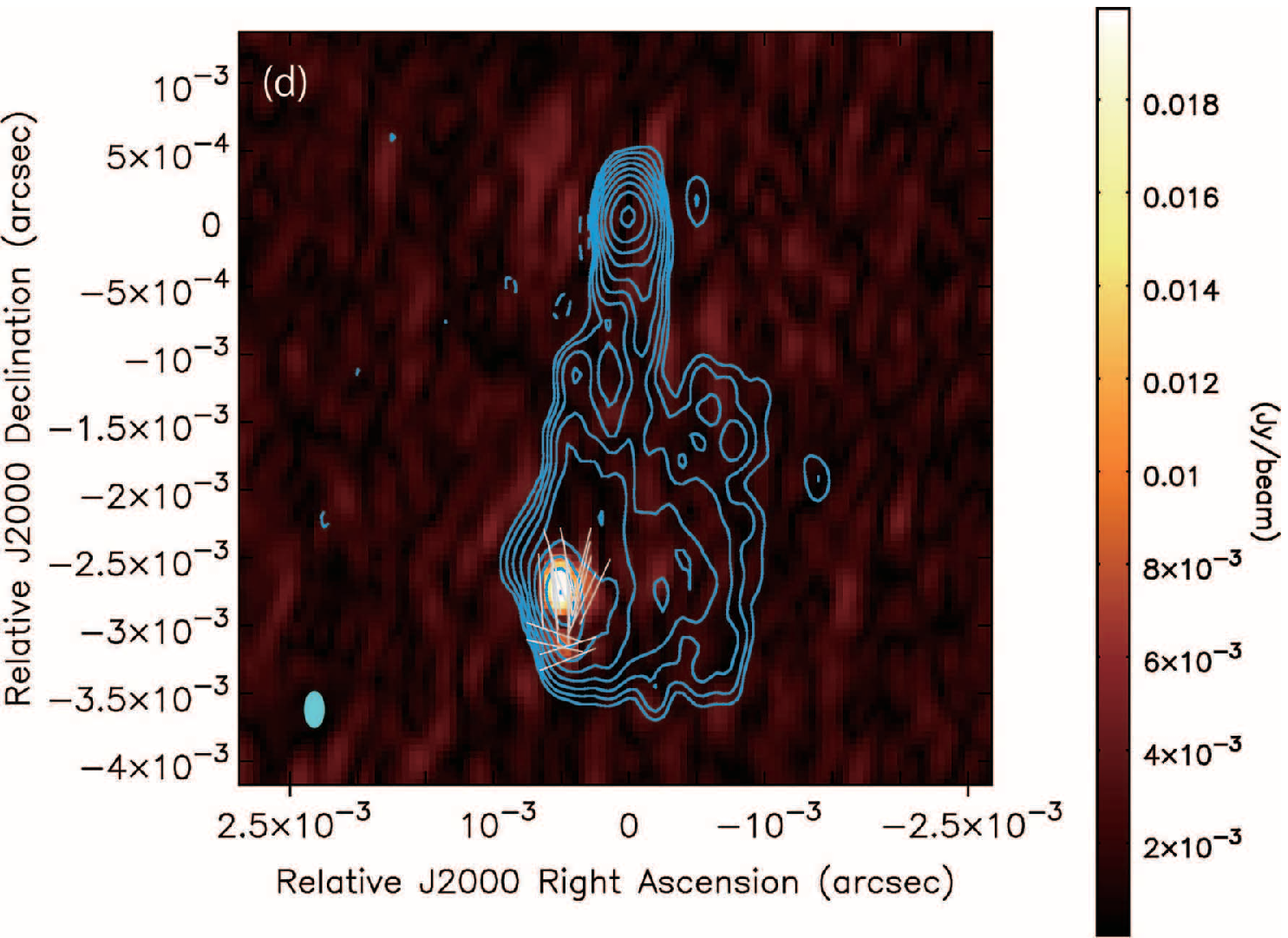}
  \end{center}
 \end{minipage}\\
\begin{minipage}{0.5\hsize}
  \begin{center}
   \includegraphics[width=8cm]{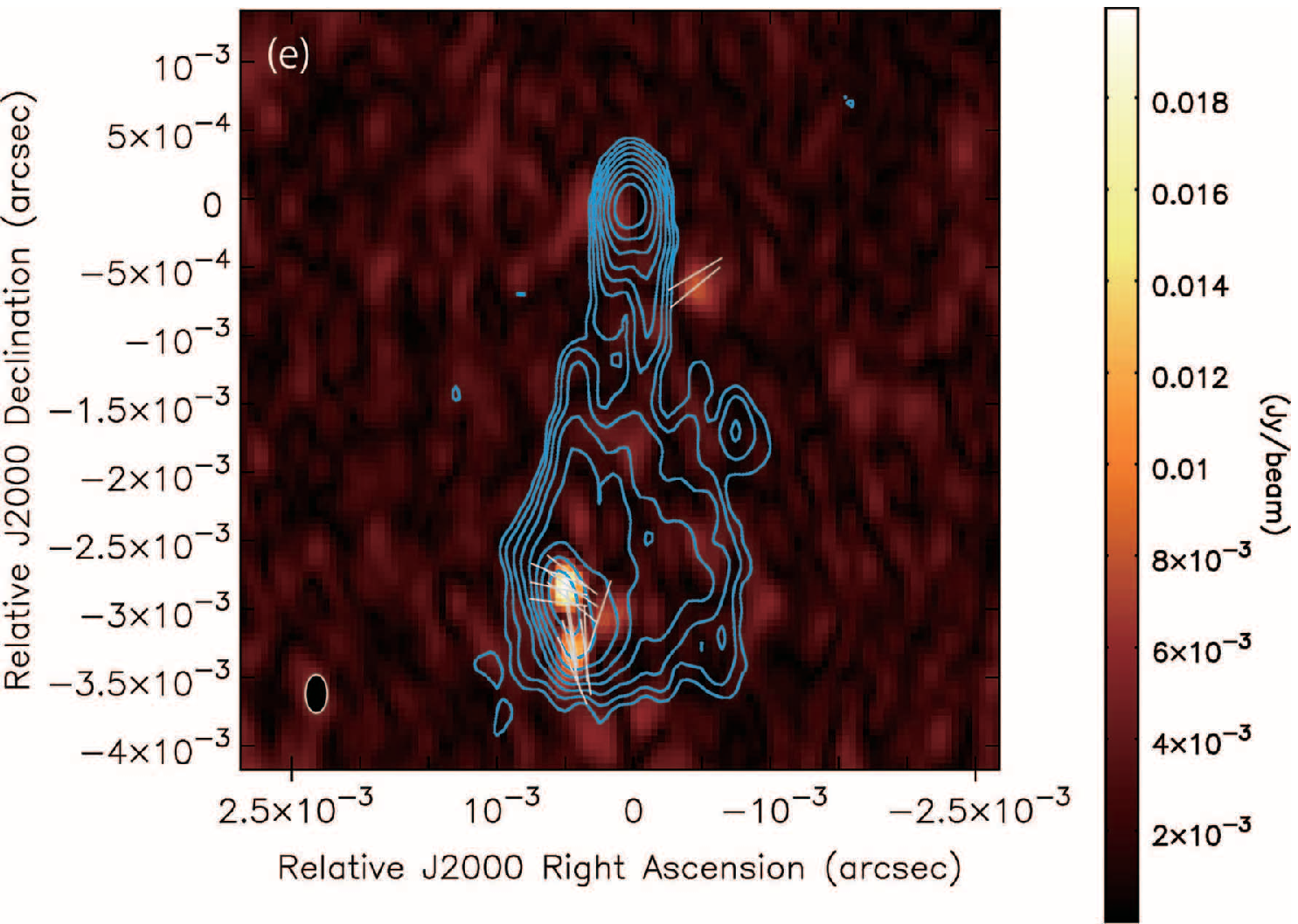}
  \end{center}
 \end{minipage}
\begin{minipage}{0.5\hsize}
  \begin{center}
   \includegraphics[width=5cm]{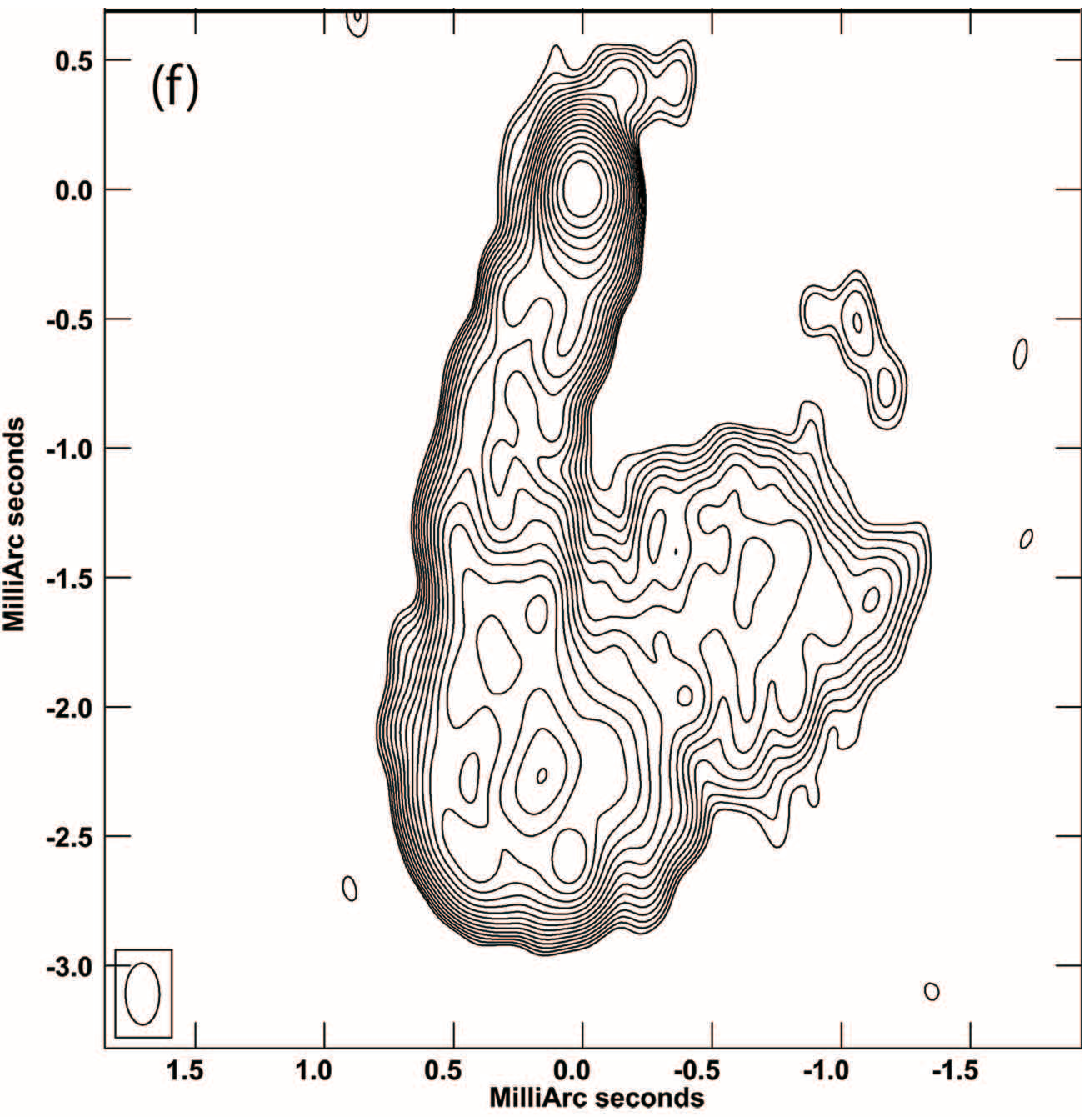}
  \end{center}
 \end{minipage}
\end{tabular}
 \caption{VLBA images of 3C~84 on (a) 2015 December 05, (b) 2016 January 01, (c) 2016 January 31, (d) 2016 March 18, and (e) 2016 April 22.  Total intensity is indicated by cyan contours overlaid on the polarized intensity in color.  Total intensity images are made by combining all four IFs while the polarized intensity is for IF1.  The 1$\sigma$ of the total intensities are 1.67~mJy, 5.56~mJy, 3.55~mJy, 3.99~mJy, and 4.44~mJy on 2015 December 05, 2016 January 01, 2016 January 31, 2016 March 18, and 2016 April 22, respectively.  The contour levels are plotted at the level of (-1, 1, 2, 4, ..., 2048) $\times$ 3$\sigma$ of each image.  The polarization position angles are indicated by the white vectors, which are shown on the area where the polarized intensity is greater than 3$\sigma$ of the polarized intensity images.  The restoring beam size is ($0.28\times0.16$)~mas at a position angle of $0^{\circ}$. (f) The 43-GHz VLBA image of 3C~84 in 2013 January 24, which is the same image shown in \cite{Nagai2014}.}
\label{fig:3C84image}
\end{center}
\end{figure*}

The imaging was done using the CLEAN algorithm implemented in Difmap software \citep{Shepherd1994}.  Each IF is imaged separately in Stokes I, Q, and U.  Figures \ref{fig:3C84image}(a)-(e) show the total and polarized intensity images. A bright and compact region, which is consistent with a hotspot, is clearly detected at the position of $\sim$3~mas from the core along the eastern limb.  In Figures \ref{fig:3C84image}(f), we also show the total intensity image in 2013 when the hotspot was seen on the western limb as a comparison.   

The peak polarized intensity, EVPA, and polarization percentage are tabulated in Table \ref{tab:Pol}.  The errors of polarized intensity and EVPA are calculated from image rms on Stokes Q and U.  

\section{RESULTS}\label{sect:results}
The polarized emission is clearly detected on the hotspot throughout all epochs at the level of 1-3\% (Table \ref{tab:Pol}).  Significance levels are about $5\sigma$ in 2015 December 05 and more than $7\sigma$ in rest of epochs.  In general, the polarization debias should be taken into account \citep{Vaillancourt2006}, but the signal-to-noise ratio is very high, and thus the debiased value of the polarized intensity is unchanged to within the uncertainty of the original value.  The main polarization feature on the hotspot is very compact and its location is not largely changed throughout these epochs.  In addition to this main polarization feature, some extended polarized emissions are detected around the hotspot in 2016 March 18 and 2016 April 22.  

\begin{figure}
\begin{center}
\includegraphics[width=7.5cm]{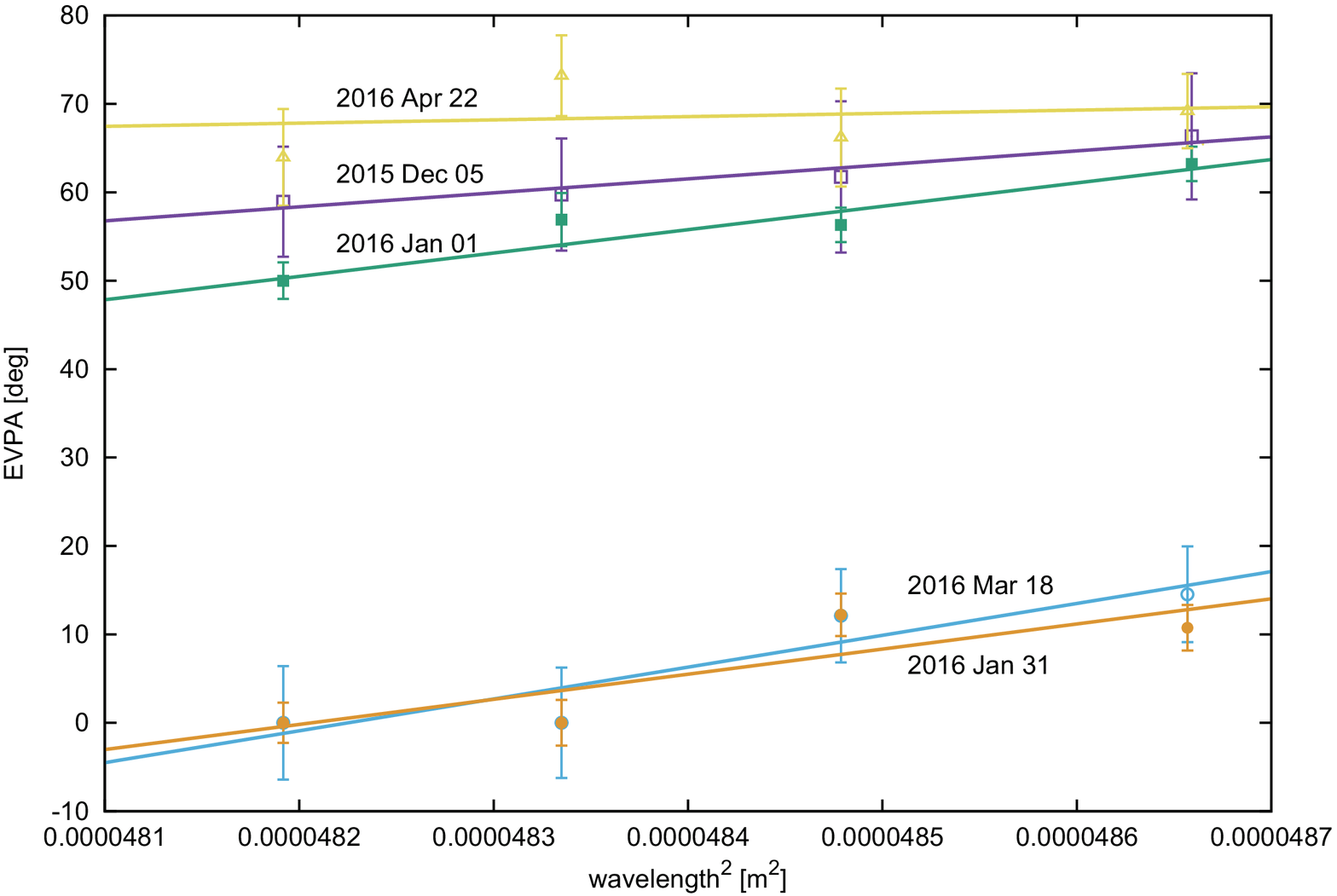}
\end{center}
\caption{EVPA as a function of wavelengths square.}
\label{fig:3C84RM}	
\end{figure}

Figure \ref{fig:3C84RM} shows the polarization position angle of the main polarized feature as a function of wavelength square.  The polarization position angle is computed from Stokes Q and U intensities at the polarized peak position.  
The rotation of EVPA with frequency is clearly detected except for data on 2016 April 22.  The inferred Faraday rotation measure (RM) is an order of 10$^{5}$ rad~m$^{-2}$ (Table \ref{tab:RM}).  This level of RM is about two orders of magnitude higher than the one reported on the polarization feature detected on the $\sim10$-mas-scale radio lobe in 2004 (Taylor et al. 2006), which is associated with the previous episode of jet activity.  The same order of RM is also reported by mm interferometric observations in 2011-2013 \citep{Plambeck2014}.   The RM might be slightly time variable (see Figure \ref{fig:3C84RM} and Table \ref{tab:RM}).     

The most intriguing finding is that the electric vector position angle (EVPA) after the correction of Faraday rotation changes abruptly on the timescale of a month (see Table \ref{tab:RM}), which is the evidence of rapid change in the projected magnetic field direction at the hotspot.  Given that the relative position of the main polarized feature to the core is almost unchanged, the change in the magnetic field direction requires that the size of the feature is smaller than 1 light month.

\begin{table}
\caption{Polarization properties of the hotspot}\label{tab:Pol}
  \begin{center}
    \begin{tabular}{ccccc} \hline\hline
Date & Freq. [GHz]	& P [mJy]\tablenotemark{a} &	$\chi$ [deg] \tablenotemark{b}	& F [\%]\tablenotemark{c}	\\ \hline
2015 Dec 05  & 43.007	& $10.2\pm1.8$ &	66.3$\pm$7.1 	& $1.28\pm0.23$	\\
& 43.087	& $9.7\pm2.1$ &	61.7$\pm$8.6 	& $1.21\pm0.26$	\\
& 43.151	& $11.2\pm1.8$ &	59.7$\pm$6.4 	& $1.40\pm0.23$	\\
& 43.215	& $11.0\pm1.8$ &	58.9$\pm$6.2 	& $1.37\pm0.22$	\\
2016 Jan 01 & 43.007	& $33.2\pm1.6$  &	63.2$\pm$2.0 	& $2.10\pm0.10$	\\
& 43.087	& $34.3\pm1.6$ & 	56.3$\pm$1.9 	& $2.21\pm0.10$	\\
& 43.151	& $31.7\pm2.1$  & 56.9$\pm$3.0 	& $2.05\pm0.14$	\\
& 43.215	& $35.6\pm1.8$  &50.0$\pm$2.1 	& $2.30\pm0.12$	\\
2016 Jan 31 & 43.008	& $15.0\pm2.0$  &	14.5$\pm$5.4 	& $1.02\pm0.14$	\\
& 43.087	& $16.8\pm2.2$  &	12.1$\pm$5.3 	& $1.14\pm0.15$	\\
& 43.151	& $15.7\pm2.3$ &	0.0$\pm$6.2 	& $1.07\pm0.16$	\\
& 43.215	& $14.1\pm2.3$ &	0.0$\pm$6.4 	& $0.86\pm0.15$	\\				
2016 Mar 18 & 43.008	& $30.8\pm2.1$ &	10.7$\pm$2.6 	& $1.08\pm0.08$	\\
& 43.087	&  $33.8\pm2.2$ &	12.2$\pm$2.4 	& $1.19\pm0.08$	\\
& 43.151	& $33.0\pm2.1$ &	0.0$\pm$2.6 	& $1.16\pm0.08$	\\
& 43.215	& $34.6\pm2.1$ &	0.0$\pm$2.3 	& $1.21\pm0.07$	\\				
2016 Apr 22 & 43.008	& $24.2\pm2.5$ &	69.2$\pm$4.2 	& $0.92\pm0.10$	\\
& 43.087	& $20.0\pm2.7$	& 66.2$\pm$5.5 	& $0.76\pm0.10$	\\
& 43.151	& $23.7\pm2.7$ &	73.2$\pm$4.6 	& $0.90\pm0.10$	\\
& 43.215	& $20.0\pm2.7$ &	64.0$\pm$5.5 	& $0.76\pm0.10$	\\
	\hline
   \end{tabular}	
  \end{center}
\tablenotetext{1}{Observed polarization position angle}
\tablenotetext{2}{Observed polarization position angle}
\tablenotetext{3}{Polarization percentage}
\end{table}

\begin{table}
\caption{Faraday rotation measure on the polarized feature.}\label{tab:RM}
  \begin{center}
    \begin{tabular}{ccc} \hline\hline
Date & Rotation measure [rad m$^{-2}$] & $\chi_{0}$ [deg]\tablenotemark{a} \\ \hline 
2015 Dec 05 & (2.7$\pm$0.5)$\times10^{5}$ & $-23\pm5$ \\
2016 Jan 01 & (4.6$\pm$0.9)$\times10^{5}$ & $89\pm19$ \\
2016 Jan 31 & (6.3$\pm$1.9)$\times10^{5}$ & $30\pm10$ \\
2016 Mar 18 & (5.0$\pm$2.2)$\times10^{5}$ & $24\pm11$ \\
2016 Apr 22 & $<2.3\times10^{5}$ &  N/A \\
\hline
   \end{tabular}	
  \end{center}
\tablenotetext{1}{Faraday-rotation-corrected EVPA}
\end{table}	

\section{DISCUSSION AND CONCLUSIONS}
\subsection{Hotspot}\label{sect:hotspot}
The polarized emission from the hotspot was not clearly detected before the middle of 2015 when the hotspot was observed on the western limb \citep[][see also the webpage of Boston University Blazar Program]{Nagai2014}.  After the hotspot appeared on the eastern limb recently, the polarized emission was suddenly enhanced.  This hotspot movement as well as the enhancement of the polarized emission presumably indicates that the jet is jittering in an inhomogeneous ambient medium, which results in the movement of the termination shock.  Similar behavior of hotspots is predicted by recent numerical simulations where the jet beam is injected in dense inhomogeneous ambient medium (e.g., Wagner et al. 2012).  Alternatively, the ejection of jet flows that represents the eastern and western limb structures could be time variable and non-simultaneous between two limbs.  

For a simple transverse shock, the magnetic field becomes perpendicular to the jet with an enhancement of fractional polarization due to the shock compression (Laing 1980).  However, the Faraday-rotation-corrected position angle indicates that the projected magnetic field is not always perpendicular to the jet axis (see Table \ref{tab:RM}).  There is no strong correlation between the fractional polarization and the project magnetic field direction.  Thus, the source of polarized emission in the hotspot of 3C~84 is presumably not only dominated by a simple transverse shock.    

Variations in the projected magnetic field are seen on the timescale of a month, which constrains the cross-sectional size of the hotspot to be less than one light month ($\sim0.07$~mas).  This upper limit of the size can be roughly interpreted as the width of the eastern jet limb since the hotspot is observed on the eastern limb.  The jet limb is not well resolved even in the recent space VLBI image with the Radioastron (beam size $\sim0.05$~mas, Giovannini et al. in preparation), which seems to be consistent with this picture.    

\subsection{Origin of Faraday Screen}\label{sect:RMorigin}
Faraday rotation can occur in any magnetized plasma along the line of sight between the source and the observer in general.  However, we can naturally expect that the observed Faraday rotation originates close to the nucleus of 3C~84, as discussed in \cite{Plambeck2014}).  Taylor et al. reported the RM of 7000 rad~m$^{-2}$ on the southern jet/lobe separated from the core by $\sim8$~pc.  Our measured RM on the hotspot separated from the core by $\sim1$~pc is at most two orders of magnitude higher than this, and thus the Faraday screen, which is responsible for the RM of $\sim10^{5}$~rad~m$^{-2}$, should originates in 1~pc$<r<$8~pc in the projected distance ($r$ is the distance from the core).  We note that the projection effect is not very significant in the estimation of distance since the jet angle to the line of sight is moderate ($65\pm16^{\circ}$: Fujita \& Nagai 2017).  The jet angle is also independently estimated by the modeling of the broadband spectrum from radio to $\gamma$-ray using synchrotron self-Compton (SSC) with some external photons, which favors smaller angles (e.g., 18$^{\circ}$: Tavecchio et al. 2014).  If this is the case, the hotspot distance is larger by a factor of a few.

The bolometric luminosity of 3C~84 is about 0.4\% of the Eddington luminosity.  Thus, the accretion flow of 3C~84 is likely to be a radiatively inefficient accretion flow (RIAF: Narayan \& Yi 2005) rather than a standard disk \citep{Shakura1973}.  We however note that 3C~84 have a cold ($T_{e}\sim10^{4}$~K) disk-like accretion flow, as identified by FFA of the emission from the counter jet in pc scale \citep{Walker2000} and inhomogeneous gas distribution around the black hole \citep{Fujita2016}.  A number of theoretical studies predicted that the accretion flow components of hot geometrically thick (RIAF-like) and cold geometrically thin can co-exist in either horizontal or vertical stratification \citep[e.g.,][]{Miller2000, Merloni2002, Liu2007, Ho2008, Liu2013}.  
The measured Faraday rotation can be caused by such a RIAF-like component.  We thus estimate the accretion rate of the RIAF-like component using the measured RM.  For a simplicity, we assume that the RIAF-like component is quasi-spherical Bondi accretion flow with a power-law density profile.  We can calculate the accretion rate, following the formulation as follows \citep{Quataert2000, Marrone2006, Kuo2014}  
\begin{eqnarray*}
\dot{M}&=&1.3\times10^{-10}\left[1-(r_{out}/r_{in})^{-(3\beta-1)/2}\right]^{-2/3}\\
&\times&\left(\frac{M_{\rm BH}}{8.0\times10^{8}M_{\odot}}\right)^{4/3}\left(\frac{2}{3\beta-1}\right)^{-2/3}r_{\rm in}^{7/6}\left(\frac{\rm RM}{\rm rad~m^{-2}}\right)^{2/3}.
\end{eqnarray*}
For inner effective radius $r_{\rm in}$ of 1~pc ($1.3\times10^{4}R_{s}$) where the hotspot is located, the observed RM implies an accretion rate of $\sim4.3\times10^{-2}M_{\odot}$~yr$^{-1}$ and $\sim8.6\times10^{-2}M_{\odot}$~yr$^{-1}$ for $\beta=0.5$ and $\beta=1.5$ which are corresponding to convection-dominated accretion flow \citep[CDAF:][]{Narayan2000, Quataert2000} and advection-dominated accretion flow \citep[ADAF:][]{Ichimaru1977,Narayan1995}, respectively.  Here we assumed the outer effective radius $r_{\rm out}$ of $10^{5} R_{s}$ ($\sim8$~pc), which is approximately the same with the Bondi radius of 8.6~pc \citep{Fujita2016}.  The derived accretion rate is roughly consistent with that estimated from the bolometric luminosity with a black hole mass of $8\times10^{8}M_{\odot}$ and a radiative efficiency of 10\% ($\dot{M}\sim L_{\rm bol}/(0.1c^{2})\simeq7.1\times10^{-2}M_{\odot}$~yr$^{-1}$). 

\begin{figure*}
\begin{center}
\includegraphics[width=15cm]{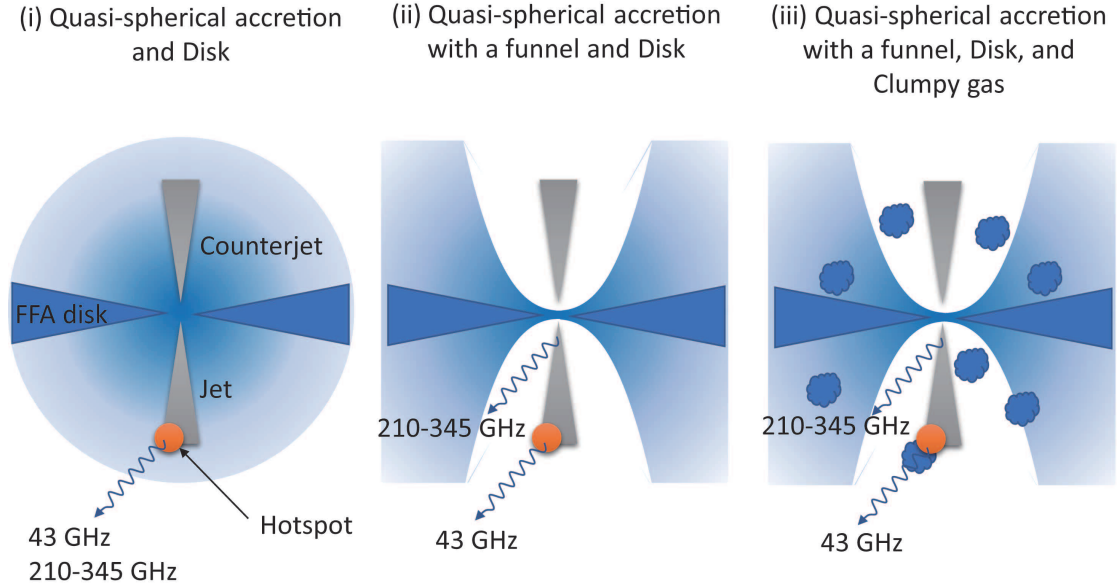}
\end{center}
\caption{Schematic image of possible accretion flow structures.  (i) 210-345~GHz emission is dominant in the hotspot.  Faraday rotation is caused by RIAF.  (ii) 210-345~GHz emission is dominant in the inner jet.  Funnel zones are created by the previous jet activities, which causes a decrease of Faraday rotation as compared to the case (i).  (iii) The jet is interacting with a clumpy gas, which responsible for the observed Faraday rotation toward the hotspot at 43~GHz.}
\label{fig:schematicview}
\end{figure*}

\cite{Plambeck2014} also reported the detection of the polarized emission at 210-345~GHz with the CARMA and SMA.  The RM is measured to be $\sim9\times10^{5}$~rad~m$^{-2}$, which agrees with the largest RM in our measurements ($6\times10^{5}$~rad~m$^{-2}$) within a factor of two.  They also estimated the accretion rate using the same method mentioned above and found that the derived accretion rate is much smaller than the one expected from the bolometric luminosity of 3C~84.  In this calculation, they assumed that the polarized emission originates in a small region at the vicinity of the black hole since the emission from the inner jet can be dominant in shorter wavelengths \citep[e.g.][]{Lobanov1998, Hada2011, Sokolovsky2011}.  One possible explanation is that the emission does not dominate at the inner jet but at the hotspot, even at 210-345~GHz (see Figure \ref{fig:schematicview} (i)).  However, one problem with this hypothesis is that the polarized flux measured by Plambeck et al. is too large to be attributable to the emission from the hotspot.  They reported about 1.5\% of the polarization percentage in 2011-2013.  Given that the total flux was 6-8~Jy in this period, the polarization flux was estimated to be about 100~mJy.  On the other hand, our measured integrated polarized flux is at most 30~mJy, which seems to be difficult to connect these two data sets unless the hotspot spectrum is inverted.  Thus, the polarized emission detected with the CARMA and SMA seems to originate in a place different from the one from the hotspot.  Consequently, a simple quasi-spherical accretion flow (Figure \ref{fig:schematicview}) cannot be applied to NGC~1275.     

The underprediction of RM by the CARMA and SMA observations indicates the necessity of reducing the accretion rate within 1~pc.   One possibility is that a part of the accretion flow is blown out by the jet and hollow ``funnels" are formed parallel to the jet axis (see Figure \ref{fig:schematicview} (ii)).  Since 3C~84 shows intermittent jet activities, the jet of previous activity can blow out the gas of the accretion flow.  The jets of ongoing activity are probably passing through the inflating ``cocoon" filled by the shocked jet material \citep{Begelman1989} of the previous jet activity.  Such a cocoon can be identified as emissions at lower frequencies \citep{Taylor1996, Silver1998, Walker2000}, which indicates that a certain amount of relativistic plasma fills the cocoon.  Faraday rotation is weakened in relativistic plasma by a factor of $\ln{\gamma/2\gamma^{2}}$, where $\gamma$ is electron Lorentz factor \citep{Quataert2000}.  Thus, we may ignore the effect of Faraday rotation within the cocoon.  
The cocoon may also explain the observed jet collimation.  \cite{Nagai2014} found a rather cylindrical collimation of the jet on subpc scale while the power-law density profile of the ambient medium, which is expected from RIAF models, predicts a jet collimation of parabolic profile \citep[e.g.][]{Narayan2011}.   The observed collimation profile can be maintained by a strong pressure of the hot cocoon.

With this configuration of accretion flow (Figure \ref{fig:schematicview} (ii)), however, the RM observed toward the hotspot can also be smaller than the case for the simple spherical accretion flow (Figure \ref{fig:schematicview} (i)).  We need an additional component to account for the observed RM toward the hotspot.  The observed Faraday rotation may be caused not only by the RIAF-like accretion flow but the dense gas localized at the hotspot (hereafter we call the gas clump).  The observed RM toward the hotspot can be preferentially larger if the line of sight to the hotspot intercepts the gas clump (see Figure \ref{fig:schematicview} (iii)).  As a result, the RM difference between the hotspot and the inner jet can be smaller.  We consider that Figure \ref{fig:schematicview} (iii) is the most likely scenario to explain the observed RM in NGC~1275 with the CARMA, SMA, and VLBA.  
This idea is also supported by the movement of the hotspot, which is probably caused by the inhomogeneous ambient medium, as we discussed in section \ref{sect:hotspot}.  Thus, we consider that the gas clump at the vicinity of the hotspot is the most likely scenario to explain the RM observed with the VLBA.  We note that such a clumpy/inhomogeneous ambient medium is also favored by the evidence of the intermittent jet activity \citep{Asada2006, Nagai2010}.  Recent numerical simulations showed that the accretion flow can be unstable because of the thermal instability in the strong X-ray field by the central source and forms cold clumps and filamentary structures on subpc-pc scales \citep{Barai2012, Gaspari2013}.  As a result, the accretion flow within pc scale cannot be spherical inflow.  Although the magnitude of RMs is lower than this value, relatively high RM is also detected in a small area of jets in a few other radio galaxies \citep[e.g.][]{Gomez2000, Zavala2002}, and attributed to the dense ionized gas in the vicinity of the radio jet.  

Here we discuss the properties of this dense gas clump.  Since the hotspot is a quite compact feature on the eastern limb, the interaction between the jet and gas clump should take place in a small area of the jet cross section. Thus, the clump size would be an order of the limb width, say 0.1~mas (0.034~pc).  \cite{Fujita2017} discussed the density of the surrounding medium in the direction of the jet based on the argument of the momentum balance along the jet.  The derived electron density ($n_{e}$) is 8.1~cm$^{-3}$ on pc scales.  To account for the observed RM of $\sim6\times10^{5}$ rad~m$^{-2}$ with this electron density, the ambient medium is required to have the magnetic field ($B$) of 2.7~G, which is much higher than the equipartition magnetic field ($B_{\rm eq}=4(\pi n_{e}k_{\rm B}T)^{0.5}$ where $k$ is Boltzmann constant) of $\sim24$~$\mu$G, with the path length of 0.034~pc.  Instead, the magnetic field becomes close to the equipartition if we adopt $n_{e}=2\times10^{4}$~cm$^{-3}$ ($B\sim B_{\rm eq}=1.2$~mG).  Thus, we conclude that the gas clump has $n_{e}$ of three orders of magnitude higher than the mean $n_{e}$ on pc scales.  Such a $n_{e}$ is consistent with the cloud of narrow emission line region around the central engine \citep{Osterbrock1991}.  It is possible that the gas clump is accreting to the SMBH since it is located within the Bondi radius of 8.6~pc \citep{Fujita2016}.  

As we showed in section \ref{sect:results}, the observed RM looks slightly time variable.  This could indicate that the gas density is inhomogeneous even within a clump, which causes temporal variations of RM as the hotspot moves.  Alternatively, the time variability of RM could be caused by the accretion flow itself.  \cite{Pang2011} predicted temporal RM variations due to the dynamical fluctuations of electron density and magnetic field in the accretion flow on the timescale of weeks to years by numerical simulations.  They suggested that the RM variations can be a tool to distinguish the accretion flow models.  The predominant component of rapid RM variations, however, is the electrons in the vicinity of radius where the electron becomes relativistic regime ($r_{rel}$).  The radius $r_{rel}$ is expected to be an order of 100$R_{s}$ \citep{Yuan2003} or even smaller \citep[see discussion by][]{Kuo2014}.  Since our measured RM originates in a region much further away, the accretion flow may not be responsible for the observed RM variations.  Further observations of RM variations will be valuable to constrain the origin of RM. 
	
\subsection{Implications to Magnetic Field Configuration}
The energy transport and dissipation mechanism from AGN to cluster scale is the primary subject for the heating of intra-cluster medium (ICM) against the radiative cooling at the cluster core \citep[e.g.][]{McNamara2007}.  Many different heating mechanisms such as the dissipation of sound waves and weak shocks \citep{Fabian2003, Fabian2005}, magnetohydrodynamic (MHD) waves \citep{Fujita2007}, cosmic rays \citep{Fujita2012, Fujita2013}, and the mixing of hot bubble gas with the ICM \citep{Hillel2017} have been proposed.  The magnetic field at the cluster core is crucial for these arguments since the energy transportation can depend on the global configuration of the magnetic field.  It is worth noting that all RMs reported in the present paper, \cite{Taylor2006}, and \cite{Plambeck2014} show the plus sign, which indicates that the mean line-of-sight magnetic field component points to the observer at three different distances from the central black hole.  This allows us to speculate that the magnetic field configuration in the ambient medium is radial from the black hole to $\sim8$~pc scale.  The RM toward the counter jet component would show minus sign if this is the case.  Measurements of the RM toward the counter jet are key to test this speculation.  A counter jet component N1 reported in \cite{Fujita2017} is possibly a hotspot interacting with the ambient medium.  A relatively ordered magnetic field is likely to exist in such a region.  Thus, the polarized emission can be detected from this component by higher-frequency VLBI observations, which can avoid a strong free-free absorption and Faraday depolarization by the accretion disk \citep[e.g.][]{Walker2000}.

As we discussed in section \ref{sect:RMorigin}, the equipartition magnetic field on pc scales is estimated to be about $24~\mu$G.  If the magnetic field configuration is radial ($B(r)\propto r^{-2}$), the magnetic field strength should be $1.6\times10^{4}$~G at $\sim1R_{S}$.  We note that the accretion flow can be inhomogeneous (see discussion in section \ref{sect:RMorigin}), which is likely to cause changes in magnetic field direction.  Thus, this is the first-order approximation.    There is no measurement of the magnetic field strength at the vicinity of the SMBH for 3C~84, so it is not possible to do a comparison of the derived value with a different approach.  One relevant information is that \cite{Kino2015} derived the magnetic field strength of about 100~G at the base of the jet of M87 using the turnover frequency due to the synchrotron self-absorption with the aid of Event Horizon Telescope (EHT) data.  This value is somewhat smaller than the case of 3C~84, but the magnetic field of 3C~84 can be much larger than M87 since the radio luminosity of 3C~84 is roughly two orders of magnitude larger than that of M87.  \cite{O'Sullivan2009} estimated the magnetic field strength at the jet base for BL Lac objects using the core-shifts argument and derived an order of $10^{4}$~G, which is similar to our estimate for 3C~84.  It seems that the radial configuration of magnetic field in the ambient medium is not too extreme assumption 

\subsection{Implications to the Origin of Low Polarization in 3C~84}
Traditionally 3C~84 is used as an instrumental polarization calibrator for most Very Large Array (VLA) frequencies and configurations because no strong polarized emission is detected.  In the image presented in this paper, the polarized emission is detected as a spot while the emissions from the remaining regions are indeed unpolarized.  This might indicate that the magnetic field of the ambient medium is highly turbulent within the VLBA beam size ($\sim0.3$~mas$\simeq0.1$~pc), resulting in a strong depolarization, and the magnetic field is somehow ordered only in the vicinity of the hotspot.  If this is the case, the jet may not be interacting with a `clumpf but a stretched or filamentary gas feature supported by the ordered magnetic field.  Such a gas feature could be a scaled-down version of the H$\alpha$ filaments observed in kpc to 10~kpc scales \citep{Fabian2008}.  However, we cannot exclude a possibility that the emissions from the jet of 3C~84 are intrinsically less polarized.  It is noteworthy that not only 3C~84 but also other radio galaxies (M87, Cen~A, Cyg~A) show no or little polarization \citep{Middelberg2005} in contrast to strongly polarized jets of blazars.  Interestingly, all these radio galaxies show a limb-brightened structure (M87: Kovalev et al. 2007; Ly et al. 2007; Hada et al. 2011, Cen~A: Kataoka et al. 2006, Cyg~A: Boccardi et al. 2016), which is suggestive of the velocity gradient across the jet width.   The number of studies on the spectral energy distribution (SED) of AGN jets also support this model \citep[e.g.][]{Ghisellini2005, Tavecchio2008, Tavecchio2014}.  The observed difference in the polarized emission between blazars and radio galaxies can be explained if the spine emission is strongly polarized while the sheath emission is less polarized.  This scenario can be testable with millimeter and submillimeter VLBI observations with the Global Millimeter VLBI Array (GMVA) and the EHT with the Atacama Large Millimeter/submillimeter Array (Array), which are less affected by the depolarization thanks to small beam and high frequency.

\bigskip 
We thank an anonymous referee for helpful comments.  This study makes use of 43~GHz VLBA data from the VLBA-BU Blazar Monitoring Program (VLBA-BU-BLAZAR; \url{http://www.bu.edu/blazars/VLBAproject.html}), funded by NASA through the Fermi Guest Investigator Program.  HN acknowledges S. Jorstad for providing the detail of calibrations for BU data and useful discussions. The VLBA is an instrument of the National Radio Astronomy Observatory. The National Radio Astronomy Observatory is a facility of the National Science Foundation operated by Associated Universities, Inc. HN is supported by MEXT KAKENHI Grant Number 15K17619.  YF is supported by MEXT KAKENHI Grant Number 15K05080.

\renewcommand{\bibname}{}


\begin{thebibliography}{}
\bibitem[Asada et al.(2006)]{Asada2006} Asada, K., Kameno, S., 
Shen, Z.-Q., et al.\ 2006, \pasj, 58, 261
\bibitem[Barai et al.(2012)]{Barai2012} Barai, P., Proga, D., \& Nagamine, K.\ 2012, \mnras, 424, 728 
\bibitem[Begelman \& Cioffi(1989)]{Begelman1989} Begelman, M.~C., \& Cioffi, D.~F.\ 1989, \apjl, 345, L21 
\bibitem[Boccardi et al.(2016)]{Boccardi2016} Boccardi, B., Krichbaum, T.~P., Bach, U., Bremer, M., \& Zensus, J.~A.\ 2016, \aap, 588, L9 
\bibitem[Dutson et al.(2014)]{Dutson2014} Dutson, K.~L., Edge, A.~C., Hinton, J.~A., et al.\ 2014, \mnras, 442, 2048  
\bibitem[Fabian et al.(2015)]{Fabian2015} Fabian, A.~C., Walker, S.~A., Pinto, C., Russell, H.~R., \& Edge, A.~C.\ 2015, \mnras, 451, 3061 
\bibitem[Fabian et al.(2008)]{Fabian2008} Fabian, A.~C., Johnstone, R.~M., Sanders, J.~S., et al.\ 2008, \nat, 454, 968 
\bibitem[Fabian et al.(2005)]{Fabian2005} Fabian, A.~C., Reynolds, C.~S., Taylor, G.~B., \& Dunn, R.~J.~H.\ 2005, \mnras, 363, 891 
\bibitem[Fabian et al.(2003)]{Fabian2003} Fabian, A.~C., Sanders, J.~S., Allen, S.~W., et al.\ 2003, \mnras, 344, L43 
\bibitem[Fanaroff \& Riley(1974)]{Fanaroff1974} Fanaroff, B.~L., \& Riley, J.~M.\ 1974, \mnras, 167, 31P 
\bibitem[Fujita \& Nagai(2017)]{Fujita2017} Fujita, Y., \& Nagai, H.\ 2017, \mnras, 465, L94 
\bibitem[Fujita et al.(2016)]{Fujita2016} Fujita, Y., Kawakatu, N., Shlosman, I., \& Ito, H.\ 2016, \mnras, 455, 2289 
\bibitem[Fujita \& Ohira(2013)]{Fujita2013} Fujita, Y., \& Ohira, Y.\ 2013, \mnras, 428, 599 
\bibitem[Fujita \& Ohira(2012)]{Fujita2012} Fujita, Y., \& Ohira, Y.\ 2012, \apj, 746, 53 
\bibitem[Fujita et al.(2007)]{Fujita2007} Fujita, Y., Suzuki, T.~K., Kudoh, T., \& Yokoyama, T.\ 2007, \apjl, 659, L1 
\bibitem[Gaspari et al.(2013)]{Gaspari2013} Gaspari, M., Ruszkowski, M., \& Oh, S.~P.\ 2013, \mnras, 432, 3401 
\bibitem[Ghisellini et al.(2005)]{Ghisellini2005} Ghisellini, G., Tavecchio, F., \& Chiaberge, M.\ 2005, \aap, 432, 401 
\bibitem[G{\'o}mez et al.(2000)]{Gomez2000} G{\'o}mez, J.-L., Marscher, A.~P., Alberdi, A., Jorstad, S.~G., \& Garc{\'{\i}}a-Mir{\'o}, C.\ 2000, Science, 289, 2317 
\bibitem[G{\'o}mez et al.(2002)]{Gomez2002} G{\'o}mez, J.-L., Marscher, A.~P., Alberdi, A., Jorstad, S.~G., \& Agudo, I., VLBA Scientific Memo No. 30
\bibitem[Hada et al.(2011)]{Hada2011} Hada, K., Doi, A., Kino, M., et al.\ 2011, \nat, 477, 185 
\bibitem[Hillel \& Soker(2017)]{Hillel2017} Hillel, S., \& Soker, N.\ 2017, \mnras, 466, L39 
\bibitem[Ho(2008)]{Ho2008} Ho, L.~C.\ 2008, \araa, 46, 475 
\bibitem[Huang \& Shcherbakov(2011)]{Huang2011} Huang, L., \& Shcherbakov, R.~V.\ 2011, \mnras, 416, 2574 
\bibitem[Ichimaru(1977)]{Ichimaru1977} Ichimaru, S.\ 1977, \apj, 214, 840 
\bibitem[Inogamov \& Sunyaev(2010)]{Inogamov2010} Inogamov, N.~A., \& Sunyaev, R.~A.\ 2010, Astronomy Letters, 36, 835 
\bibitem[Jorstad et al.(2005)]{Jorstad2005} Jorstad, S.~G., Marscher, A.~P., Lister, M.~L., et al.\ 2005, \aj, 130, 1418 
\bibitem[Kataoka et al.(2006)]{Kataoka2006} Kataoka, J., Stawarz, {\L}., Aharonian, F., et al.\ 2006, \apj, 641, 158 
\bibitem[Kino et al.(2015)]{Kino2015} Kino, M., Takahara, F., Hada, K., et al.\ 2015, \apj, 803, 30
\bibitem[Kovalev et al.(2007)]{Kovalev2007} Kovalev, Y.~Y., Lister, M.~L., Homan, D.~C., \& Kellermann, K.~I.\ 2007, \apjl, 668, L27 
\bibitem[Kuo et al.(2014)]{Kuo2014} Kuo, C.~Y., Asada, K., Rao, R., et al.\ 2014, \apjl, 783, L33 
\bibitem[Lim et al.(2008)]{Lim2008} Lim, J., Ao, Y., \& Dinh-V-Trung 2008, \apj, 672, 252-265 
\bibitem[Liu et al.(2007)]{Liu2007} Liu, B.~F., Taam, R.~E., Meyer-Hofmeister, E., \& Meyer, F.\ 2007, \apj, 671, 695 
\bibitem[Liu \& Taam(2013)]{Liu2013} Liu, B.~F., \& Taam, R.~E.\ 2013, \apjs, 207, 17 
\bibitem[Lobanov(1998)]{Lobanov1998} Lobanov, A.~P.\ 1998, \aaps, 132, 261 
\bibitem[Ly et al.(2007)]{Ly2007} Ly, C., Walker, R.~C., \& Junor, W.\ 2007, \apj, 660, 200 
\bibitem[Marrone et al.(2006)]{Marrone2006} Marrone, D.~P., Moran, J.~M., Zhao, J.-H., \& Rao, R.\ 2006, \apj, 640, 308 
\bibitem[McNamara \& Nulsen(2007)]{McNamara2007} McNamara, B.~R., \& Nulsen, P.~E.~J.\ 2007, \araa, 45, 117
\bibitem[Merloni \& Fabian(2002)]{Merloni2002} Merloni, A., \& Fabian, A.~C.\ 2002, \mnras, 332, 165  
\bibitem[Melrose(1997)]{Melrose1997} Melrose, D.~B.\ 1997, Journal of Plasma Physics, 57, 479 
\bibitem[Middelberg et al.(2005)]{Middelberg2005} Middelberg, E., Roy, A.~L., Bach, U., Gabuzda, D.~C., \& Beckert, T.\ 2005, Future Directions in High Resolution Astronomy, 340, 189 
\bibitem[Miller \& Stone(2000)]{Miller2000} Miller, K.~A., \& Stone, J.~M.\ 2000, \apj, 534, 398 
\bibitem[Nagai et al.(2010)]{Nagai2010} Nagai, H., Suzuki, K., 
Asada, K., et al.\ 2010, \pasj, 62, L11  
\bibitem[Nagai et al.(2014)]{Nagai2014} Nagai, H., Haga, T., Giovannini, G., et al.\ 2014, \apj, 785, 53 
\bibitem[Narayan \& Yi(1995)]{Narayan1995} Narayan, R., \& Yi, I.\ 1995, \apj, 444, 231 
\bibitem[Narayan et al.(2000)]{Narayan2000} Narayan, R., Igumenshchev, I.~V., \& Abramowicz, M.~A.\ 2000, \apj, 539, 798 
\bibitem[Narayan \& Fabian(2011)]{Narayan2011} Narayan, R., \& Fabian, A.~C.\ 2011, \mnras, 415, 3721 
\bibitem[Osterbrock(1991)]{Osterbrock1991} Osterbrock, D.~E.\ 1991, Reports on Progress in Physics, 54, 579 
\bibitem[O'Sullivan \& Gabuzda(2009)]{O'Sullivan2009} O'Sullivan, S.~P., \& Gabuzda, D.~C.\ 2009, \mnras, 400, 26 
\bibitem[Pang et al.(2011)]{Pang2011} Pang, B., Pen, U.-L., Matzner, C.~D., Green, S.~R., \& Liebend{\"o}rfer, M.\ 2011, \mnras, 415, 1228 
\bibitem[Perlman et al.(2007)]{Perlman2007} Perlman, E.~S., Mason, R.~E., Packham, C., et al.\ 2007, \apj, 663, 808 
\bibitem[Plambeck et al.(2014)]{Plambeck2014} Plambeck, R.~L., Bower, G.~C., Rao, R., et al.\ 2014, \apj, 797, 66  
\bibitem[Quataert \& Gruzinov(2000)]{Quataert2000} Quataert, E., \& Gruzinov, A.\ 2000, \apj, 545, 842 
\bibitem[Romney et al.(1995)]{Romney1995} Romney, J.~D., Benson, 
J.~M., Dhawan, V., et al.\ 1995, Proceedings of the National Academy of 
Science, 92, 11360 
\bibitem[Salom{\'e} et al.(2006)]{Salome2006} Salom{\'e}, P., Combes, F., Edge, A.~C., et al.\ 2006, \aap, 454, 437 
\bibitem[Scharw{\"a}chter et al.(2013)]{Scharwachter2013} Scharw{\"a}chter, J., McGregor, P.~J., Dopita, M.~A., \& Beck, T.~L.\ 2013, \mnras, 429, 2315 
\bibitem[Shakura \& Sunyaev(1973)]{Shakura1973} Shakura, N.~I., \& Sunyaev, R.~A.\ 1973, \aap, 24, 337 
\bibitem[Shepherd et al.(1994)]{Shepherd1994} Shepherd, M.~C., 
Pearson, T.~J., \& Taylor, G.~B.\ 1994, \baas, 26, 987 
\bibitem[Silver et al.(1998)]{Silver1998} Silver, C.~S., Taylor, G.~B., \& Vermeulen, R.~C.\ 1998, \apj, 502, 229 
\bibitem[Sokolovsky et al.(2011)]{Sokolovsky2011} Sokolovsky, K.~V., Kovalev, Y.~Y., Pushkarev, A.~B., \& Lobanov, A.~P.\ 2011, \aap, 532, A38 
\bibitem[Sosa-Brito et al.(2001)]{Sosa-Brito2001} Sosa-Brito, R.~M., Tacconi-Garman, L.~E., Lehnert, M.~D., \& Gallimore, J.~F.\ 2001, \apjs, 136, 61
\bibitem[Taylor \& Vermeulen(1996)]{Taylor1996} Taylor, G.~B., \& Vermeulen, R.~C.\ 1996, \apjl, 457, L69 
\bibitem[Taylor et al.(2006)]{Taylor2006} Taylor, G.~B., Gugliucci, N.~E., Fabian, A.~C., et al.\ 2006, \mnras, 368, 1500 
\bibitem[Tavecchio \& Ghisellini(2014)]{Tavecchio2014} Tavecchio, F., \& Ghisellini, G.\ 2014, \mnras, 443, 1224 
\bibitem[Tavecchio \& Ghisellini(2008)]{Tavecchio2008} Tavecchio, F., \& Ghisellini, G.\ 2008, \mnras, 385, L98 
\bibitem[Tan et al.(2008)]{Tan2008} Tan, J.~C., Beuther, H., Walter, F., \& Blackman, E.~G.\ 2008, \apj, 689, 775-781 
\bibitem[Vaillancourt(2006)]{Vaillancourt2006} Vaillancourt, J.~E.\ 2006, \pasp, 118, 1340 
\bibitem[Wagner et al.(2012)]{Wagner2012} Wagner, A.~Y., Bicknell, G.~V., \& Umemura, M.\ 2012, \apj, 757, 136 
\bibitem[Walker et al.(2000)]{Walker2000} Walker, R.~C., Dhawan, 
V., Romney, J.~D., Kellermann, K.~I., 
\& Vermeulen, R.~C.\ 2000, \apj, 530, 233 
\bibitem[Yuan et al.(2003)]{Yuan2003} Yuan, F., Quataert, E., \& Narayan, R.\ 2003, \apj, 598, 301 
\bibitem[Zavala \& Taylor(2002)]{Zavala2002} Zavala, R.~T., \& Taylor, G.~B.\ 2002, \apjl, 566, L9 


\end{thebibliography}
\end{document}